\documentclass[11pt,a4paper]{iopart}

\usepackage{iopams}    % need for subequations
\usepackage{graphicx}   % need for figures
\usepackage{color}      % use if color is used in text

\begin{document}

\title{Two-photon absorption in gapped bilayer graphene with a tunable 
chemical potential}
\author{M.~K.~Brinkley$^1$}
\author{D.~S.~L.~Abergel$^{2,3}$}
\author{B.~D.~Clader$^1$}
\address{$^1$Johns Hopkins University Applied Physics Laboratory, Research and Exploratory Development Department, Laurel, MD 20973}
\address{$^2$Nordita, KTH Royal Institute of Technology and Stockholm University, Roslagstullsbacken 23, SE-106 91 Stockholm, Sweden}
\address{$^3$Center for Quantum Materials, KTH and Nordita,
Roslagstullsbacken 11, SE-106 91 Stockholm, Sweden}

%\pacs{}

\begin{abstract}
Despite the now vast body of two-dimensional materials under study, bilayer graphene remains unique in two ways:  it hosts a simultaneously tunable band gap and electron density; and stems from simple fabrication methods. These two advantages underscore why bilayer graphene is critical as a material for optoelectronic applications. In the work that follows, we calculate the one- and two-photon absorption coefficients for degenerate interband absorption in a graphene bilayer hosting an asymmetry gap and adjustable chemical potential---all at finite temperature. Our analysis is comprehensive, characterizing one- and two-photon absorptive behavior over wide ranges of photon energy, gap, chemical potential, and thermal broadening. The two-photon absorption coefficient for bilayer graphene displays a rich structure as a function of photon energy and band gap due to the existence of multiple absorption pathways and the nontrivial dispersion of the low energy bands. This systematic work will prove integral to the design of bilayer-graphene-based nonlinear optical devices.

\end{abstract}

\maketitle

The excitement following the isolation of graphene \cite{abergel_review}
is due in part to the remarkable optoelectronic properties the
material possesses
\cite{geim1,geim2}. In addition to a high electron
mobility \cite{mobility} and fast optical response \cite{ultrafast},
graphene and its bilayer (BLG) exhibit tunable broadband optical
absorption \cite{geim2,abergel_falko,abergel_das_sarma,modulator1,exciton},
suggesting applicability to such devices as optical modulators
\cite{modulator1, modulator2},
photodetectors \cite{photodetector1,photodetector2}, and possibly
all-optical switches \cite{switch1,switch2,switch3,switch4}. Recent
studies have revealed that graphene has a large third-order
susceptibility \cite{four-wave1,four-wave2}, leading to a strong
two-photon absorption (2PA) coefficient \cite{photonics_review},
which is stronger yet in the bilayer system \cite{giant2PA} due to the
nested manifold of bands present at the $K$-point of the Brillouin zone.
As a result of this $K$-point band-commensuration and the larger number
of absorptive pathways, the 2PA coefficient for the ungapped bilayer
system is several orders of magnitude higher than that of monolayer
graphene (MLG) in certain frequency ranges \cite{giant2PA}.

Despite the attractive optoelectronic properties of undoped and gapless 
graphene systems, the absence of a gap at the $K$-point precludes
MLG from use in many device applications. Multilayer graphene
systems, however, do exhibit a band gap when chemically doped
or externally gated, permitting a wide range of control over the
conductivity.
In fact, by simultaneous use of top and bottom gating, the carrier
density and gap size can be independently tuned \cite{henriksen}, 
providing a degree of dynamical control of the optical response not
available in many optoelectronic materials. 
Below, we demonstrate that the 2PA of bilayer graphene displays a rich
structure as a function of these two parameters.

Moreover, as the doping concentration of graphene is very
sensitive \cite{fab_doping} to synthesis methodology and conditions, a
characterization of doped, gapped graphene systems will prove paramount
to experimental characterization and successful device design.

At present, a theoretical characterization of the 2PA strength in doped,
gapped BLG has yet to be reported, and calculations of the 2PA
coefficient \cite{giant2PA} in the ungapped system are incomplete since
they do not encompass all
possible intermediate states. Here, we compute using a perturbative
approach the full one- and two-photon absorption coefficients for a
graphene bilayer with a band gap and a tunable chemical potential. The
physical scenario we discuss is a back-gated graphene bilayer placed
underneath a transparent top gate (see, for example,
Ref.~\cite{bolometer}), providing simultaneous and independent control
over the chemical potential, $\mu$, and the asymmetry gap, $\Delta$. 
The one-photon absorption (1PA) spectrum for the gapped system has been
computed previously by Nicol and Carbotte \cite{carbotte}, which we
reproduce for comparison to the 2PA spectrum. 

\begin{figure*}[t]
	\centering
	\includegraphics[width=\textwidth]{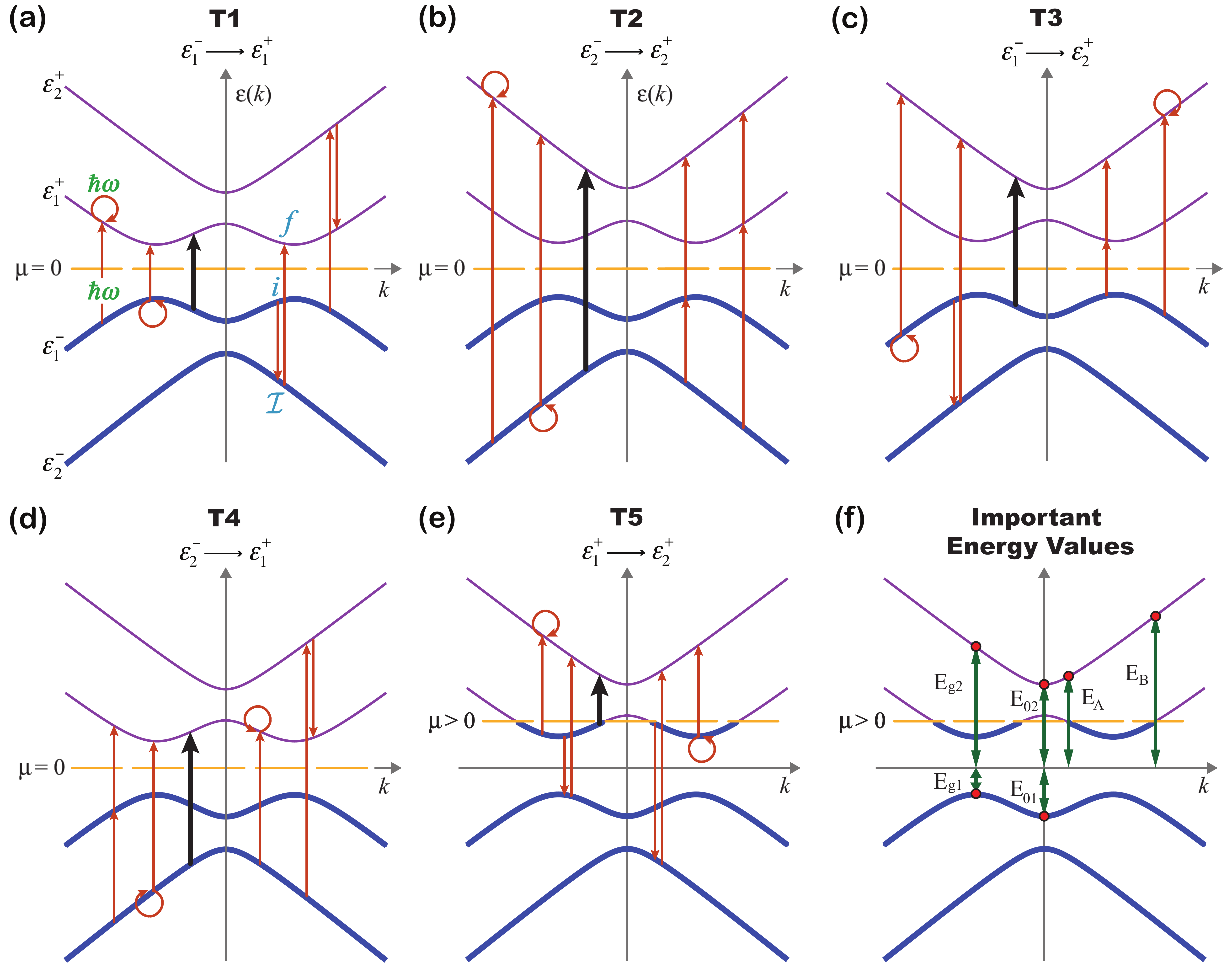}
	\caption{\label{fig:Fig1} (a)-(e) Illustrations of the allowed 1PA
	(thick {black} arrows) and 2PA (thin {red} arrows) transitions,
	$T1-T5$, in a gapped graphene bilayer at $T=0$. The thick dashed
	line represents the chemical potential, and the thick (thin) curves
	denote populated (unpopulated) bands. The circular arrows signify
	transitions for which $|i\rangle \longrightarrow
	|\mathcal{I}=i\rangle$ or $|\mathcal{I}\rangle \longrightarrow
	|f=\mathcal{I}\rangle$. (f) Important energy values for the gapped
	band structure are indicated with green arrows, as defined in 
	Ref.~\cite{carbotte}. }
\end{figure*}

An electric field oriented perpendicular to a Bernal-stacked graphene
bilayer gives rise to an asymmetry gap \cite{mccann}, for which the
tight-binding Hamiltonian in the $K$-valley using the basis
(A1,B2,A2,B1) in the sublattice space is
\begin{equation} 
	\hat{\mathcal{H}}_{TB} = 
	\left[ \begin{array}{cccc} 
		-\Delta/2 & 0 & 0 & \gamma_1 \rho {\rm e}^{-i\phi} \\ 
		0 & \Delta/2 & \gamma_1 \rho {\rm e}^{i\phi} & 0 \\ 
		0 & \gamma_1 \rho {\rm e}^{-i\phi} & \Delta/2 & \gamma_1 \\ 
		\gamma_1 \rho {\rm e}^{i\phi} & 0 & \gamma_1 & -\Delta/2
	\end{array} \right]. 
\end{equation}
We use the following notation: $\phi = \tan^{-1}(p_y/p_x)$ is the polar
angle of the in-plane momentum, $\textbf{p}$; $\rho \equiv v_F p /
\gamma_1$ is the reduced momentum; $\gamma_1 = 0.4$ eV is the A2-B1
interlayer coupling; and $v_F \sim10^6\mathrm{ms}^{-1}$ is the Fermi
velocity. 
The corresponding exact eigenvalues and eigenvectors are 
\begin{equation} 
	\varepsilon^{\pm}_j = \pm \left( \gamma_1^2 \rho^2 
		+ \frac{\Delta^2}{4} + \frac{\gamma_1^2}{2} + 
		\gamma_1 (-1)^j \sqrt{\rho^2 (\Delta^2 + \gamma_1^2) 
		+ \frac{\gamma_1^2}{4}} \right)^{{1}/{2}}
\end{equation}
and 
\begin{equation} 
	|\psi^{\pm}_j(\rho)\rangle = \frac{1}{\mathcal{N}_j} 
	\left[\begin{array}{c}
	\frac{2\gamma_1 \rho \exp{(-i\phi)}}{2 \varepsilon^{\pm}_j+\Delta} \\
	\frac{2\gamma_1^2 \rho \exp{(i\phi)}}{\mathcal{C}_j 
		- 2 \varepsilon^{\pm}_j \Delta} \\  
	\frac{\gamma_1 (2\varepsilon^{\pm}_2-\Delta)}{\mathcal{C}_j 
		- 2 \varepsilon^{\pm}_j \Delta} \\
	1
	\end{array}\right]
\end{equation}
where $j = 1(2)$ denotes the low-energy (high-energy) bands split by
$\Delta$ at the $K$-point; $\mathcal{N}_j$ is a normalization factor;
and $\mathcal{C}_j \equiv \Delta^2 + \gamma_1^2 + (-1)^j \sqrt{4
\Delta^2 \rho^2 + 4 \gamma_1^2 \rho^2 + \gamma_1^2}$. For an incident
field with polarization chosen along the $x$ direction, ${\bf A}(t)= A_0
\hat{x} \exp(i\omega t)$, the interaction Hamiltonian %\cite{gauge} 
is
\begin{equation} 
	\hat{\mathcal{H}}_{int} =   \frac{e}{2mc} {\bf A}  \cdot \hat{\bf p} 
	=  \frac{e}{2c} \left| \frac{i c}{\omega} {\bf E} \right| 
	{\hat x} \cdot \hat{\bf v} = \frac{e}{2\omega} E_0 
	\frac {\partial \hat{\mathcal{H}}_{TB}} {\partial {p_x}}. 
\end{equation} 
Assuming a relative permittivity $\epsilon_r \approx 9$ for
BLG \cite{epsilonr,epsilonr_note} and an incident irradiance $I=E_0^2 c
\epsilon_0 \sqrt{\epsilon_r}/2$, perturbation theory gives the one- and
two-photon absorption coefficients \cite{nathan,vanstryland}, which are,
respectively, 
\begin{equation} 
	\label{1OP}
	\beta_1 =\frac{2 \hbar \omega}{I} \frac{2\pi}{\hbar} g \sum_{i,f} 
	\left| \langle f |\hat{\mathcal{H}}_{int} |i\rangle \right|^2
	\delta (\varepsilon_{f}-\varepsilon_{i} - \hbar\omega)
\end{equation} 
and
\begin{equation}
	\label{2OP}
	\beta_2 =  \frac{4 \hbar \omega}{I^2} \frac{2\pi}{\hbar} g \sum_{i,f} 
	\left| \sum_{\mathcal I} \frac {\langle f | \hat{\mathcal{H}}_{int} |
	{\mathcal I} \rangle \langle {\mathcal I} | \hat{\mathcal{H}}_{int} 
	|i\rangle}{\varepsilon_{\mathcal I}-\varepsilon_{i}
	-\hbar \omega} \right|^2
	\delta
	(\varepsilon_{f}-\varepsilon_{i}-2\hbar\omega),  
\end{equation} 
where $\beta_n \equiv 2n \hbar\omega W_n / I^n$ is the $n$-photon
absorption coefficient; $W_n$ represents the $n^{\mathrm{th}}$-order 
interband transition probability rate per unit area; $g=4$ is a factor
accounting for spin and valley degeneracy; and the subscripts label the
initial ($i$), intermediate ($\mathcal{I}$) and final ($f$) states.

\begin{figure*}[t]
	\centering
	\includegraphics[width=\textwidth]{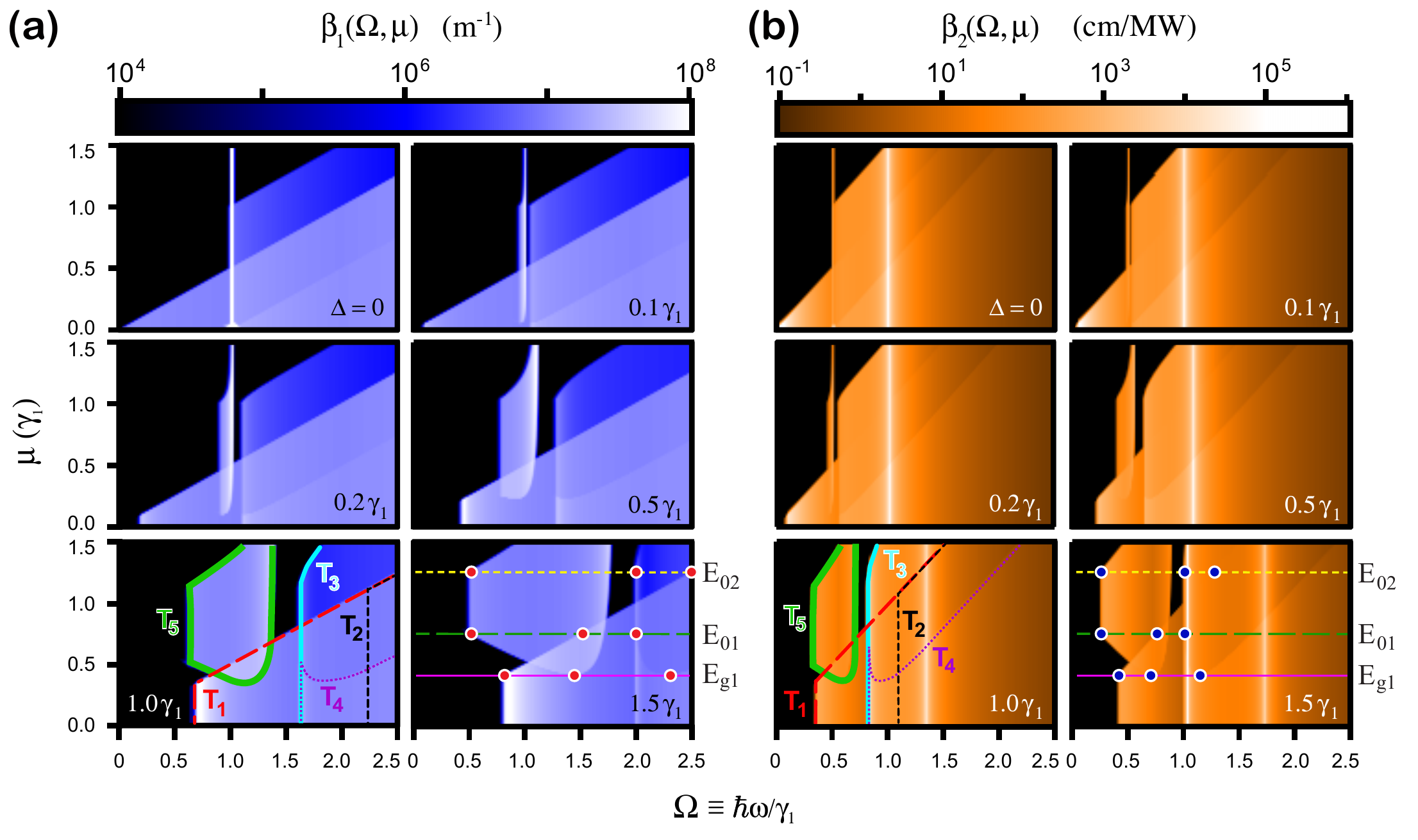}
	\caption{\label{fig:Fig2} Color plots of the (a) one-photon absorption
	coefficient, $\beta_1(\Omega,\mu)$, and (b) two-photon absorption
	coefficient, $\beta_2(\Omega,\mu)$, for a wide range of the gap,
	$\Delta$, and chemical potential, $\mu$. The color scale at the top of
	each panel indicates the intensity of $\beta_1$ and $\beta_2$.  The
	($\mu$,$\Omega$) regions of spectral weight associated with each
	transition, T1-T5, are delineated in both (a) and (b) for
	$\Delta=\gamma_1$, showing how each transition (labeled in
	Fig.~\ref{fig:Fig1})
	contributes to the 1PA or 2PA coefficient. 
	The panels for $\Delta = 1.5 \gamma_1$ depict the relevant energy
	thresholds illustrated in Fig.~\ref{fig:Fig1}(f).  
	The colored circles mark the changes in spectral weight that result
	as $\mu$ crosses $E_{g1}$, $E_{01}$, and $E_{02}$.} 
\end{figure*}

Figure \ref{fig:Fig1} shows the band structure at the $K$-point for a
gapped graphene bilayer. The interband 1PA and 2PA transitions are
shown, each of which must satisfy the constraints of energy and momentum
conservation imposed by the $\delta$-functions in the expressions for
$\beta_n$. For the case of 1PA, the thick black arrows in
Fig.~\ref{fig:Fig1}(a)-(d)
denote the four one-photon pathways (T1-T4) permitted from
$\varepsilon_i \longrightarrow \varepsilon_f$ for a neutral chemical
potential ($\mu=0$) at a temperature of $T=0$ K.  The thick arrow in the
fifth panel, Fig.~\ref{fig:Fig1}(e), illustrates T5, the transition allowed from
$\varepsilon^{+}_1 \longrightarrow \varepsilon^{+}_2$ when $\mu>0$. 

Associated with each 1PA transition is a family of possible 2PA pathways
\cite{rioux}, which are illustrated in Fig.~\ref{fig:Fig1} by thin red
arrows. 
In the case of 2PA, two photons are absorbed simultaneously via two
energy-nonconserving transitions of $|i\rangle \longrightarrow
|\mathcal{I}\rangle$ and $|\mathcal{I}\rangle \longrightarrow
|f\rangle$, for which energy is conserved over $|i\rangle
\longrightarrow |f\rangle$. 
For each of the initial  and final state combinations, four pathways are
possible, all of which are indicated in Fig.~\ref{fig:Fig1}. The
transitions T3 and T4 displayed in Fig.~\ref{fig:Fig1}(c) and (d) are
degenerate. 
A determination of the criteria imposed on T1-T5 by energy
conservation requires the calculation of several important energy
values, which are provided for reference in Fig.~\ref{fig:Fig1}(f) using the
nomenclature of Nicol and Carbotte \cite{carbotte} and we shall refer to
these values below. 
We confine our analysis to the case $\mu \geq 0$ since electron-hole
symmetry implies that the response is identical for $\mu < 0$.

The diagrams in Fig.~\ref{fig:Fig1} and the denominator of
Eq.~(\ref{2OP}) allow for several immediate insights into the frequency
response of $\beta_2$ when $\Delta = 0$ and $T=0$.
The transitions taking part in the 2PA process for which $|i\rangle
\longrightarrow |\mathcal{I}=i\rangle$ or $|\mathcal{I}\rangle
\longrightarrow |f=\mathcal{I}\rangle$, denoted by circular arrows,
produce a singularity in $\beta_2$ at $\hbar \omega = 0$. When $\Delta$
becomes large compared to the temperature, this resonance disappears
since there are no pairs of initial and final states which satisfy the
$\delta$-function. Similarly,
contributions for which $\varepsilon_f-\varepsilon_i = 2\hbar\omega=
2\gamma_1$ produce a singularity at $\gamma_1$,
and 2PA transitions for which $\varepsilon_f-\varepsilon_i = 2
\hbar\omega= \gamma_1$ give rise to a singularity at $\hbar \omega =
\gamma_1 / 2$. For 1PA when $\mu\ge 0$ and $T>0$, a
singularity exists only due to T5, which generates an intense, narrow
peak at $\hbar\omega = \gamma_1$ when bands $\varepsilon^+_1$ and
$\varepsilon^+_2$ are perfectly nested. This peak will exist for $T>0$
even when $\mu = 0$ due to  thermal population of conduction band
$\varepsilon^{+}_1(\rho)$ and depopulation of valence band
$\varepsilon^{-}_1(\rho)$.

\begin{figure}[t]
	\centering
	\includegraphics[scale=0.52]{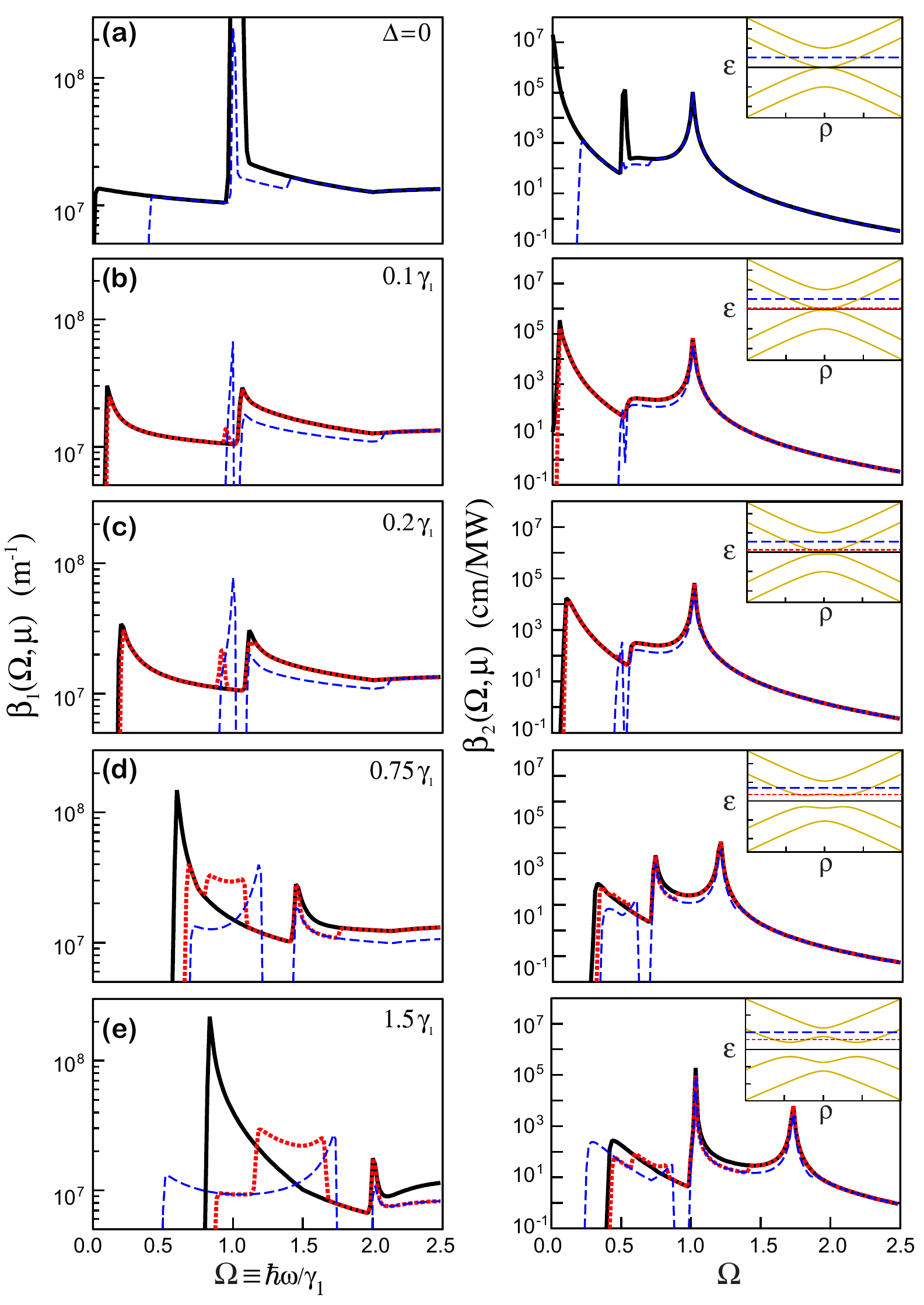}
	\caption{\label{fig:Fig3} 1PA (left column) and 2PA (right column)
	for the indicated value of the gap, $\Delta$. The solid, dotted, and
	dashed curves correspond to the identical lines superimposed on the
	dispersion curves in the insets in the right-hand column, indicating
	the absorption for chemical potentials of $\mu=0$ (continuous black
	line), $\mu=(E_{g1}+E_{01})/2$ (dotted red line), and
	$\mu=(E_{01}+E_{02})/2$ (dashed blue line).} 
\end{figure}

When an asymmetry gap $\Delta > 0$ is present, the conditions imposed by
energy conservation become more complex \cite{carbotte} due to the
so-called ``sombrero'' structure \cite{mccann} of the low-energy bands
$\varepsilon^{\pm}_1$. The bands are not perfectly nested as they are for
$\Delta=0$, leading to the rich absorptive structure displayed in
Fig.~\ref{fig:Fig2}. Figure \ref{fig:Fig2} shows color plots of both the
1PA and 2PA coefficients for a wide range of gap values.
Note that the highest value of $\Delta=1.5$ is probably unattainable
using current experimental techniques, but we wish to display the
effects when $\Delta \geq 1$ since several spectral features change.
A transformation of $\omega \longrightarrow \omega
+ i \Gamma/\hbar$ accounts for thermal broadening and we take $\Gamma =
5.4 \times 10^{-3} \gamma_1$, which corresponds to a temperature of
$T=25$ K. 

The thermal broadening leads to the emergence in Fig.~\ref{fig:Fig2}(a)
of an intense, narrow peak at $\hbar\omega=\gamma_1$ due to T5 for 
$\Delta = 0$ (see Ref.~\cite{Delta0note}). At $T = 25$ K, band
$\varepsilon^+_1$ receives spectral weight due to thermal population,
giving rise to T5 transitions of $\varepsilon^+_1
\longrightarrow\varepsilon^+_2$.
Similarly, in the $\beta_2$ spectrum for $\Delta = 0$ in
Fig.~\ref{fig:Fig1}(b), the thermal smearing of $\mu$ leads to a peak at
$\hbar\omega = \gamma_1/2$, as anticipated during the analysis of
Fig.~\ref{fig:Fig1}. 
When $\Delta$ is small, the $1/\Omega$ divergence in the 2PA is caused
by the resonance near the center of the Brillouin zone. This is not
present in the 1PA because it is cancelled by the absorption matrix
element.
When $\Delta$ increases, the T5 component of both
$\beta_1$ and $\beta_2$ broadens into the cleaver-shaped regions
outlined in green for $\Delta = \gamma_1$ in Fig.~\ref{fig:Fig2}(a,b).
The asymmetric broadening of the T5 region is due to the sombrero shape
of $\varepsilon^{\pm}_1$ and the loss of spectral weight when $\mu >
E_{02}$.  The panels corresponding to $\Delta = \gamma_1$ in
Fig.~\ref{fig:Fig2}(a) and (b) identify the energy thresholds mapped in
Fig.~\ref{fig:Fig1}(f), illustrating regions where the chemical
potential either blocks or permits absorptive pathways. For $\Delta>0$,
the band gap means that there is no absorption at $\Omega<2E_{g1}$ and
the increased density of states at the band edge produces regions of
increasing 1PA and 2PA near the T1 cutoff. The other spectrally empty
region that emerges with increasing $\Delta$ occurs between the T3/T4
and T5 pockets. 

In contrast with the $\varepsilon_f-\varepsilon_i = \hbar \omega$
requirement of 1PA, 2PA requires $\varepsilon_f=\varepsilon_i = 2\hbar
\omega$, which compresses and shifts the T1-T5 regions in $\beta_2$
relative to those $\beta_1$. 
Also, Eq.~(\ref{2OP}) shows that the resonances will be stronger in the
2PA due to the additional factor of $(\hbar\omega-\varepsilon_{\mathcal{I}}
+ \varepsilon_i)^{-2}$ in the sum over intermediate states.
The spectra for $\beta_2$ shown in Fig.~\ref{fig:Fig2}(b) contain a
prominent resonance not present in $\beta_1$. 
This $\beta_2$ resonance, arising at {$\Omega=1$} when $\Delta=0$, is
due to transitions T3 and T4, which contain denominators
of $\hbar \omega - \gamma_1$ and give rise to a singularity. As $\Delta$
increases, this resonance shifts toward higher $\Omega$ due to the
increasing size of the band gap. At this T3/T4 resonance, the BLG 2PA
strength is several orders of magnitude larger for bilayer graphene than
for monolayer graphene \cite{giant2PA}.

\begin{figure*}[t]
	\centering
	\includegraphics[scale=0.9]{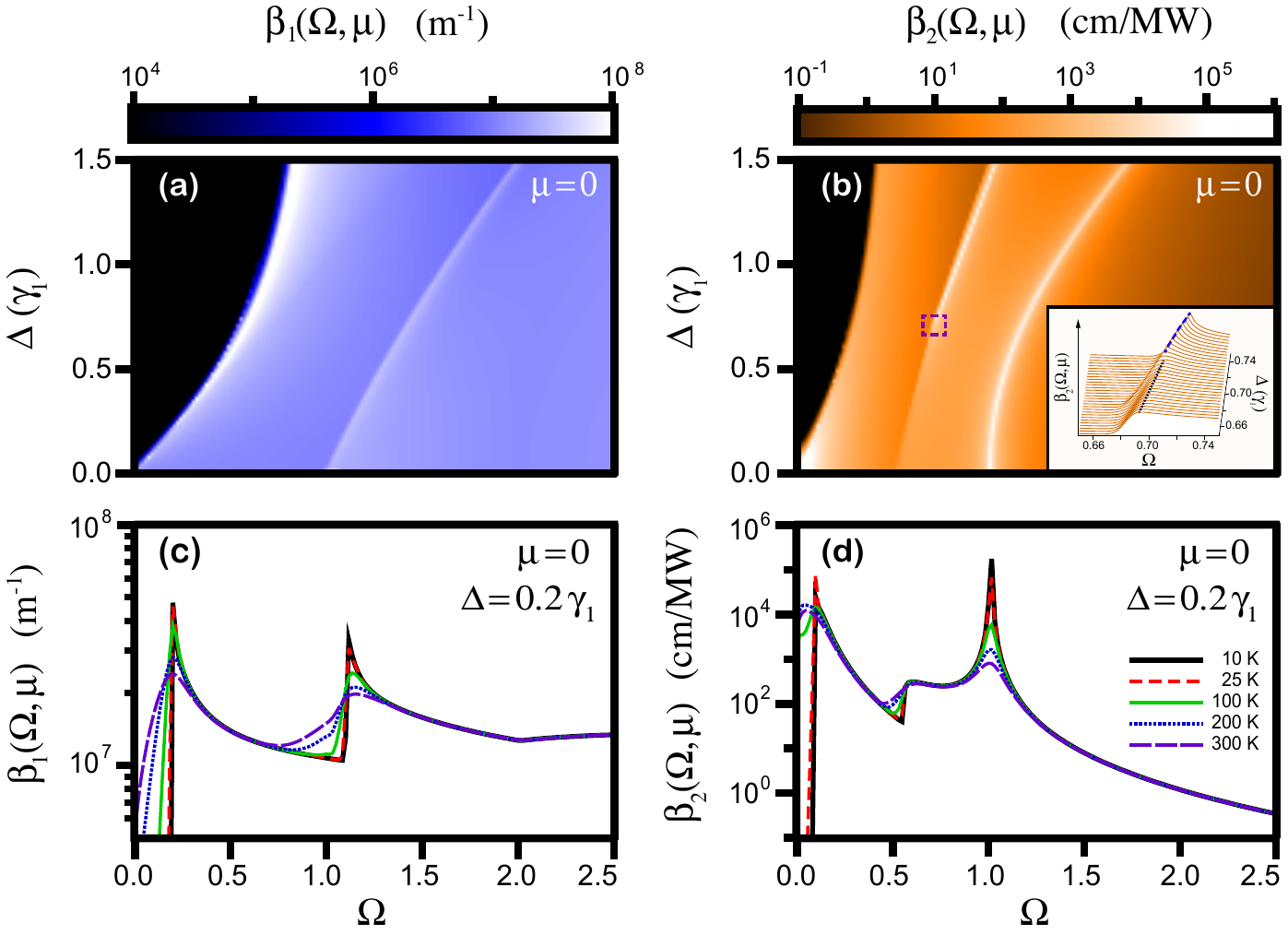}
	\caption{\label{fig:Fig4}Spectra of the (a) one-photon absorption
	coefficient, $\beta_1(\Omega,\Delta)$, and (b) two-photon absorption
	coefficient, $\beta_2(\Omega,\Delta)$, for which $\mu=0$ and
	$\Gamma$ corresponds to $25$ K. The inset in (b) contains the region
	outlined by the dashed purple box, in which a dashed blue line
	traces an emerging peak and a black dotted line indicates the T3/T4
	cutoff. Panels (c) and (d) display spectra of $\beta_1(\Omega,T)$
	and $\beta_2(\Omega,T)$, respectively, for $\Delta = 0.2 \gamma_1$
	and $\mu=0$.}
\end{figure*}

Figure \ref{fig:Fig3} presents 1PA and 2PA curves extracted from the color
maps of Fig.~\ref{fig:Fig2}, grouped by gap value. 
The left-hand column shows $\beta_1(\Omega,\mu)$, and the right-hand
column shows $\beta_2(\Omega,\mu)$. 
For each plot of $\beta_1$ and $\beta_2$, several values of
$\mu$ are chosen, each of which corresponds to an identical horizontal
line on the accompanying dispersion shown in the inset to the $\beta_2$
plots.
The solid black curves are for $\mu =0$; the dotted red  is for
$\mu=(E_{g1}+E_{01})/2$ which is where the chemical potential is in the
middle of the sombrero region; the dashed blue line is for
$\mu =(E_{01}+E_{02})/2$, and lies halfway between $\varepsilon^+_1$
and $\varepsilon^+_2$ at $\rho=0$. 
When $\Delta=0$, the dotted red slice is equivalent to that of $\mu=0$,
so that the solid black and dotted red lines are identical. 
As $\Omega \rightarrow \infty$, the contributions to $\beta_2$ of
transitions T1-T4 tend to $1/\Omega^{4}$.

Figure \ref{fig:Fig3} reveals an intense, narrow peak in $\beta_2$ that
emerges when $\Delta \ge {\gamma_1}/{\sqrt{2}}$, resulting from two
$|i\rangle \rightarrow |\mathcal{I}\rangle$ routes within the family of
T3 and T4 pathways. In particular, routes $\varepsilon^{-}_1 \rightarrow
\varepsilon^{+}_1$ and $\varepsilon^{-}_1 \rightarrow \varepsilon^{-}_2$
lead to a vanishing denominator ${\varepsilon_{\mathcal I} 
- \varepsilon_{i}-\hbar \omega}$ in Eq.~(\ref{2OP}) at values of
\begin{equation} 
	\Omega_A \equiv \frac{\sqrt{2}}{4\gamma_1}
	\left(5(\Delta^2+\gamma_1^2)-3
	\sqrt{\Delta^4+\case{2}{9}\Delta^2\gamma_1^2+\gamma_1^4} \right)^{1/2}.
\end{equation} 
When $\Omega = \Omega_A$ and {$\Delta \ge { \gamma_1 }/{\sqrt{2}}$} for
$\Gamma=0$, the position of the peak, $\Omega_A$, {eclipses} the T3/T4
cutoff, causing the peak to appear when  the photon energy $\hbar\omega
= \Omega \gamma_1$ coincides with  $\varepsilon^{+}_1-\varepsilon^{-}_1
\ge (E_{01}+E_{02})/2$ and $\varepsilon^{-}_1-\varepsilon^{-}_2\ge
(E_{01}+E_{02})/2$.

Figure \ref{fig:Fig4}(a) and (b) show $\beta_1$ and $\beta_2$ as a
function of $\Omega$ and $\Delta$ for $T=25$ K and $\mu=0$. As $\Delta$
increases from $0$, certain regions of increasing intensity materialize
in both 1PA and 2PA spectra, including the peak in $\beta_2$ at
$\Omega_A$, which is outlined in Fig.~\ref{fig:Fig4}(b) by a dashed
purple box surrounding the point $(\Omega,\Delta) = (1/ \sqrt{2},
\gamma_1/ \sqrt{2})$. The inset in Fig.~\ref{fig:Fig4}(b) provides a
zoomed-in view of the boxed region, in which the dotted black line
indicates the T3/T4 cutoff, and the dashed blue line shows the onset of
the emerging peak. Thus far, the assumed thermal
broadening corresponds to a temperature of $25$ K --- a value large
enough to reveal the T5 resonance when $\mu = 0$, yet small enough to
avoid the smearing of fine features. Figure \ref{fig:Fig4} (c) and (d)
illustrate the effect of thermal broadening on $\beta_1$ and $\beta_2$
for $\Delta = 0.2\gamma_1$ at neutral doping ($\mu=0$). The most
pronounced impact of increasing temperature is the broadening of
$\beta_1$ and $\beta_2$ into
the gap region. Once $\Gamma$ reaches room temperature, the otherwise
abrupt T3/T4 cutoff just beyond $\Omega = 1$ spreads into a more diffuse
spectral bulge. Similarly, the sharp $\beta_2$ resonance residing near
$\Omega = 1$ at $T=10$ K suffers appreciably within the temperature
range examined, dropping in intensity by more than two orders of
magnitude when $T = 300$ K.  

We have demonstrated that when an electric field is applied
perpendicular to a graphene bilayer, the resulting asymmetry gap gives
rise to complex linear and nonlinear optical absorption. 
This is experimentally verifiable in a simple photoabsorption
measurement using transparent gates.
For the gapped bilayer system with a tunable chemical potential, we have
calculated both one- and two-photon absorption spectra over an expansive
range of the gap and chemical potential, taking into account all
possible absorption pathways in the calculation of
$\beta_2(\Omega,\mu)$. 
{We analyze the 2PA resonances that emerge in the gapped, doped bilayer
system, and examine the evolution of these resonances as a function of
$\Delta$ and $\mu$.}  
The effects of thermal broadening are incorporated into the
computations, providing insight into the degradation of optical
performance at or near room temperature. As graphene-based optical
architectures mature, the absorption spectra calculated above will prove
important for optimizing and enhancing device performance. 

\section*{Acknowledgments}
This work was supported by Internal Research \& Development funds and
the Stuart S. Janney Fellowship Program at the Johns Hopkins University
Applied Physics Laboratory. MKB thanks Tai-Chang Chiang, Yang Liu, Scott
Hendrickson, and Joan Hoffmann for their valuable comments and insight.
DLSA is supported by Nordita and by ERC project DM-321031.


\begin{thebibliography}{5}
\bibitem{epsilonr_note} The value of $\epsilon_r$ for BLG will vary somewhat according to the choice of substrate, which affects only the irradiance prefactor, $1/I^{n}$, in the $n$-photon absorption coefficient, $\beta_{n} $.
\bibitem{Delta0note} For plots labeled $\Delta = 0$, we employed a gap value of $\Delta = 10^{-5}\gamma_1$ in the corresponding computations.
\end{thebibliography}
\end{document}